\documentstyle[twocolumn,prl,aps]{revtex}

  \newcommand{\beq}{\begin{equation}}
  \newcommand{\eeq}{\end{equation}}
  \newcommand{\beql}[1]{\begin{equation}\label{eq:#1}}
  \newcommand{\beqa}{\begin{eqnarray}}
  \newcommand{\eeqa}{\end{eqnarray}}
  \newcommand{\beqas}{\begin{eqnarray*}}
  \newcommand{\eeqas}{\end{eqnarray*}}

  \newcommand{\bA}{{\bf A}}

  \newcommand{\bP}{{\bf P}}
  
  \newcommand{\bS}{{\bf S}}

  \newcommand{\bx}{{\bf x}}

  \newcommand{\cH}{{\cal H}}

  \newcommand{\cK}{{\cal K}}

  \newcommand{\hL}{{\hat L}}

  \newcommand{\hS}{{\hat S}}

  \newcommand{\al}{\alpha}
  \newcommand{\be}{\beta} 
  
  \newcommand{\da}{\dagger}
  
  \newcommand{\ep}{\epsilon}
  \newcommand{\et}{\eta}

  \newcommand{\nn}{\nonumber}

  \newcommand{\ps}{\psi}

  \newcommand{\De}{\Delta}                                          
  
  \newcommand{\Eq}[1]{Eq.~(\ref{eq:#1})}

  \newcommand{\eq}[1]{(\ref{eq:#1})}

\newcommand{\ket}[1]{|#1\rangle}

\newcommand{\bracket}[1]{\langle#1\rangle}

\begin{document}
\draft
\title{Conservation laws, uncertainty relations, and quantum limits of
measurements}
\author{Masanao Ozawa}
\address{Center for Photonic Communication and Computing,
Department of Electrical and Computer Engineering\\ 
Northwestern
University, 2145 Sheridan Road, Evanston, IL 60208-3118	\\ 
and\\ 
CREST, Japan Science and Technology, 
Graduate School of Information Sciences\\
Tohoku University, Aoba-ku, Sendai,
Miyagi 980-8579, Japan}

\maketitle

\begin{abstract} 
The uncertainty relation between the noise operator and
the conserved quantity leads to a bound for the accuracy of general
measurements. The bound extends the assertion by Wigner, Araki, and
Yanase that conservation laws limit the accuracy of ``repeatable'', or
``nondisturbing'', measurements to general measurements, and improves
the one previously obtained by Yanase for spin measurements.  The
bound also sets an obstacle to making a small quantum computer. 
\end{abstract}

\narrowtext

In 1952, Wigner \cite{2} found that conservation laws put a limit on
measurements of quantum mechanical observables.   In 1960, Araki and
Yanase \cite{3} proved the following assertion known as the
Wigner-Araki-Yanase (WAY) theorem:   {\em Observables which do not
commute with bounded additive conserved quantities have no ``exact''
measurements} \cite{4}.   Subsequently, Yanase \cite{5} found a bound
for the accuracy of  measurements of the $x$-component of spin in terms 
of the ``size'' of the apparatus, where the ``size'' is characterized by the
mean-square of the $z$-component of the angular momentum \cite{6}. 
Yanase \cite{5} and Wigner \cite{7} concluded from this result that in
order to increase the accuracy of spin measurement one has to use a
very large measuring  apparatus.

In the WAY theorem, for a measurement to be ``exact''
the following two conditions are required to be satisfied: (i) the Born
statistical formula (BSF) and  (ii) the repeatability hypothesis (RH),
asserting that if an observable is measured twice in succession in a
system, then we obtain the same value each time.
Yanase's bound does not assume the RH.  Instead, a condition, to be 
referred to as Yanase's condition, is assumed  that the probe observable, the
observable in the apparatus to be measured after the measuring interaction,
commutes with the conserved quantity, to ensure the measurability of the
probe observable \cite{5}. Elaborating the suggestions given by Stein and
Shimony \cite{4}, Ohira and Pearle \cite{8} constructed a simple 
measuring interaction that satisfies the conservation law and the BSF,
assuming the precise probe measurement, but does not satisfy the RH. 
Based on their model, Ohira and Pearle claimed that it is possible to have
an accurate measurement of spin component regardless of the size of the
apparatus, if the RH is abandoned.  However, their model does not satisfy
Yanase's condition, so that the problem remains as to the measurability of
the probe observable. 

Yanase's argument, however, assumes a large (but of variable size)
measuring apparatus having the continuous angular momentum 
from the beginning for technical reasons and concludes that accurate 
measurement requires a very large apparatus.    
To avoid a circular argument, a rigorous derivation without such an assumption is
still demanded.   Moreover,  Wigner \cite{7} pointed out the necessity for
generalizing the bound to general quantum systems other than spin 1/2 systems, as well as
including  all additive conservation laws.  

In order to accomplish the suggested generalization,  a new approach to
the problem is proposed in this letter based on the uncertainty relation
between the conserved quantity and the noise operator, defined as the 
difference between the post-measurement probe and the measured
quantity.  We obtain a bound for the mean-square error of general
measuring interactions imposed by any additive conservation laws
without assuming the RH.  This bound also clarifies the trade-off between
the size and the commutativity of the noise operator with the conserved
quantity, unifying the suggestion by WAY and others and
the one suggested by Ohira and Pearle.  For spin measurements, this bound
with Yanase's condition leads to a tight bound for the error probability of
spin measurement, which improves Yanase's bound.

Let $\bA(\bx)$ be a measuring apparatus with macroscopic output
variable $\bx$ to measure, possibly with some error, an observable $A$ 
of the {\em object\/} $\bS$, a quantum system represented by a Hilbert
space $\cH$.  The measuring interaction turns on at time $t$, the {\em
time of measurement}, and turns off at time $t+\De t$ between the object
$\bS$ and the {\em probe} $\bP$, a part of the apparatus that interacts
with the object, represented by a Hilbert space $\cK$.  Denote by $U$ the
unitary operator on $\cH\otimes\cK$ representing  the time evolution of
$\bS+\bP$ in the time interval $(t,t+\Delta t)$. 

At the time of measurement the object is supposed to be in an unknown
(vector) state $\psi$ and the probe is supposed to be prepared in a known
(vector) state $\xi$; all state vectors are assumed to be normalized unless
stated otherwise. Thus, the composite system $\bS+\bP$ is in the state
$\psi\otimes\xi$ at time $t$.  Just after the measuring interaction, the probe
is subjected to a local interaction with the subsequent stages of the
apparatus.  The last process is assumed to measure an observable $M$,
called the {\em probe observable}, of the probe with arbitrary precision,
and the outcome is recorded as the value of the macroscopic outcome
variable $\bx$.  

In the Heisenberg picture with the original state $\psi\otimes\xi$ at time
$t$, we shall write 
$A(t)=A\otimes I$,
$M(t)=I\otimes M$,
$A(t+\Delta t)=U^{\dagger}(A\otimes I)U$, and $M(t+\Delta
t)=U^{\dagger}(I\otimes M)U$. 
We shall denote by ``$\bx(t)\in\Delta$'' the
probabilistic event that the outcome of the measurement using apparatus
$\bA(\bx)$  at time $t$ is in an interval
$\Delta$. Since the outcome  of this measurement is obtained by the
measurement of  the probe observable $M$ at time $t+\Delta t$, 
the probability distribution of the output variable $\bx$ is given by
\beql{B1}\label{eq:0328b}
\Pr\{\bx(t)\in\De\}=\|E^{M(t+\Delta t)}(\De)(\psi\otimes\xi)\|^{2},
\eeq where $E^{M(t+\De t)}(\De)$ stands for the spectral projection of the
operator $M(t+\De t)$ corresponding to the interval $\De$.  We call
the above description of the measuring process as the {\em indirect
measurement model} determined by
$(\cK,\xi,U,M)$ \cite{indirect}.

We say that apparatus $\bA(\bx)$ {\em measures} observable $A$
{\em precisely}, if  $\bA(\bx)$ satisfies the BSF for observable
$A$, 
\beql{010530a}
\Pr\{\bx(t)\in\Delta\}=\|E^{A}(\Delta)\psi\|^{2}
\eeq 
on every input  state $\ps$.  Otherwise, we consider apparatus
$\bA(\bx)$ to measure observable $A$ with some noise. 

The {\em noise operator} $N$ of apparatus $\bA(\bx)$ for measuring
$A$ is defined by 
\begin{equation} N=M(t+\Delta t)-A(t). \label{1.1a}
\end{equation} The {\em noise} $\ep(\psi)$ of apparatus $\bA(\bx)$ for
measuring $A$ on input state $\psi$ is, then,  defined by 
\begin{equation}
\ep({\psi})^{2}=\langle N^{2}\rangle, \label{3}
\end{equation} where $\langle \cdots\rangle$ stands for 
$\langle \psi\otimes\xi|\cdots|\psi\otimes\xi\rangle$. The noise $\ep(\psi)$
represents the  root-mean-square error in the outcome of the measurement.
By Eq.~(\ref{3}), we have 
\beql{N<E}
\ep(\psi)^{2}\ge(\Delta N)^{2},
\eeq where $\Delta X$ stands for the standard deviation  of an observable
$X$ in $\psi\otimes\xi$, i.e., 
$ (\Delta X)^{2}=\langle X^{2}\rangle-\langle X\rangle^{2}.
$

We define the {\em noise} $\ep$ of apparatus $\bA(\bx)$ to be the least
upper bound of $\ep(\ps)$ over all possible input states $\ps$.   One of the
fundamental properties of the noise is that precise apparatuses and
noiseless apparatuses are equivalent notions, as ensured by the following
theorem
\cite{note1}:  {\em Apparatus $\bA(\bx)$ measures observable $A$
precisely if and only if $\ep=0$.}

Consider the additive conservation law (ACL) for observables 
$L_{1}$ of the object $\bS$ and $L_{2}$ of the probe $\bP$, i.e.,
\begin{equation}\label{eq:ACL} [U,L_1 \otimes I + I \otimes L_2] = 0. 
\label{1.2}
\end{equation} 
In the Heisenberg picture, we shall write
$L_{1}(t)=L_{1}\otimes I$,
$L_{2}(t)=I \otimes L_{2}$,
$L_{1}(t+\Delta t)=U^{\dagger}(L_{1} \otimes I)U$, and
$L_{2}(t+\Delta t)=U^{\dagger}(I \otimes L_{2})U$.
The ACL, (\ref{1.2}),  can be restated as the invariance principle
\beql{invariance} 
L_{1}(t)+L_{2}(t) =L_{1}(t+\Delta t)+L_{2}(t+\Delta t).
\eeq

The WAY theorem \cite{2,3} states that if $L_{1}$ is
bounded, there is no apparatus precisely measuring $A$ that
satisfies the RH and the ACL, unless $A$ commutes with the conserved
quantity $L_{1}$.   In the following argument, we shall require the ACL but
abandon the RH.

Why does the conservation law limit the accuracy of measurement in
general?  A simple observation on the noise operator will lead to a
significant interplay between the conservation law and the uncertainty
relation.   As we have discussed above, the measurement is precise if and
only if 
$\bracket{N^{2}}=\|N(\ps\otimes\xi)\|^{2}=0$.   If this is the case, the
uncertainty relation
\beql{uncertainty} 
(\Delta N)^{2}(\Delta [L_{1}(t)+L_{2}(t)])^{2}\ge
\frac{1}{4}|\bracket{[N,L_{1}(t)+L_{2}(t)]}|^{2}
\eeq 
concludes that if the conserved quantity does not commute with the
noise operator in the initial state, the conserved quantity should have
infinite variance, or the precise measurement is impossible for the
bounded conserved quantity.

Let us study the quantitative relations shown by the uncertainty relation,
\eq{uncertainty}, in detail.  Since $L_{1}(t)$ and $L_{2}(t)$ are
statistically independent,  the variance of their sum is the sum of their
variances, i.e.,  
\beql{independence} 
(\Delta[L_{1}(t)+L_{2}(t)])^{2} =[\Delta
L_{1}(t)]^{2}+[\Delta L_{2}(t)]^{2}.
\eeq 
Since $A$ and $L_{1}$ are in the object and $M$ and $L_{2}$ 
are in the probe, we have
$$
[M(t+\De t),L_{1}(t+\De t)] =[A(t),L_{2}(t)]=0.
$$
By the ACL, \eq{invariance}, we obtain
\begin{eqnarray}\label{eq:commutation}\label{eq:A1.3b}
\lefteqn{[N,L_{1}(t) + L_{2}(t)]}\quad\nn\\ 
&=&[M(t+\De t),L_{2}(t+\De t)]-[A(t),L_{1}(t)].
\end{eqnarray}
From Eqs.~\eq{N<E}, \eq{uncertainty},
\eq{independence}, and \eq{commutation}, 
we obtain the following fundamental lower bound of the noise of apparatus
$\bA(\bx)$.
\beql{fundamental}
\ep(\ps)^{2}
\ge\frac{|\bracket{[M(t+\De t),L_{2}(t+\De t)]-[A(t),L_{1}(t)]}|^{2}}
{4[\Delta L_{1}(t)]^{2}+4[\Delta L_{2}(t)]^{2}}.
\eeq

From the above lower bound, in order to attain $\ep=0$ it is necessary
to choose $\xi$, $U$, and $M$ satisfying 
\beq
\bracket{\xi|U^{\da}(I\otimes[M,L_{2}])U|\xi}=[A,L_{1}].
\eeq
Stein and Shimony \cite{4} and Ohira and Pearle \cite{8} gave 
examples that actually attain $\ep=0$.
Does this mean that if we abandon the RH, the ACL allows to have a
noiseless measuring apparatus regardless of the size of the apparatus?
Recall that the noise $\ep$ is defined as the one from the measuring
interaction, which quantum mechanics can analyze in detail.
Thus, if we do not assume that the probe measurement is 
carried out precisely, the noise $\ep$ depends on the boundary 
between the probe and the rest of the apparatus. 
Since this boundary is rather arbitrary,
it can be seen that the measuring apparatus carries out the precise
measurement if and only if the noise $\ep$ vanishes for any boundaries.
Thus, in order to show that the ACL limits the accuracy of the measuring
apparatus, it suffices to show that a particular boundary leads to an
inevitable noise.  For this purpose, we shall consider the maximal boundary
in a given apparatus.  In this case, the notion of the probe is identical
with a quantum mechanical description of a measuring apparatus, so that
we can assume (i) the probe includes all the external sources of
interactions, and (ii) the probe observable plays a role of the record.
Assumption (i) is justified, since the measuring apparatus operates
covariantly so that it can be used in any laboratory and at any time. 
Assumption (ii) is justified, since the measuring apparatus includes a
record which the observer can access repeatedly.  
From assumption (i) we can assume that the measuring interaction 
satisfies the ACL.  From assumption (ii) we can assume that the probe
observable can be measured by another external measuring apparatus
satisfying the RH.
Then, the WAY theorem requires that {\em the probe observable should
commute with the additive conserved quantities}; we call this condition
Yanase's condition.
Therefore, the above argument supports our hypothesis below that {\em
in  any measuring apparatus there is a boundary between the probe and
the rest of the apparatus for which the ACL and Yanase's condition hold}.

Now, we assume Yanase's condition
\begin{equation} 
[M,L_{2}] = 0.
\end{equation} 
In this case, 
the fundamental noise bound, \eq{fundamental},
turns out to be the following form.
\beql{A1.6}\label{eq:bound}
\ep(\ps)^{2}\ge\frac{|\bracket{[A(t),L_{1}(t)]}|^{2}} {4[\Delta
L_{1}(t)]^{2}+4[\Delta L_{2}(t)]^{2}}.
\eeq
Since the input state is unknown but the probe is
prepared in a known state, the bound \label{eq:A1.6} is optimized when
the input-independent quantity $\Delta L_{2}(t)$ is maximized.  Thus,
we can conclude that {\em in order to decrease the noise of the
apparatus, one has to increase the variance of the conserved quantity in
the probe.}

Consider the case where the object $\bS$ is a particle of spin 1/2.  Let
$\hat{S}_{x}$, $\hat{S}_{y}$, and $\hat{S}_{z}$ be the spin
observables of $\bS$ in the $x$, $y$, and $z$ directions, respectively; we
shall write $\al_{i}=\ket{\hS_{i}=\hbar/2}$ and
$\be_{i}=\ket{\hS_{i}=-\hbar/2}$ for $i=x,y,z$.    In what follows, we
shall optimize the noise $\ep$ of apparatus
$\bA(\bx)$ for measuring the $x$-component of  the spin of  particle
$\bS$, under the following constraints:  (i) the measuring interaction
preserves the $z$-component of the total angular momentum, i.e.,
\beq [U,\hS_{z}+\hL_{z}]=0,
\eeq  
where $\hL_{z}$ is the $z$-component of the angular momentum of 
probe $\bP$, and (ii) the probe observable $M$  commutes with the
conserved quantity, i.e., 
\beq [M,\hL_{z}]=0.
\eeq 
By the optimization it is meant, here, to obtain the saddle point in
which the bound is maximized by the object state and minimized by the
probe state.  From the above constraints, \Eq{bound} holds for
$A=\hS_{x}$, $L_{1}=\hS_{z}$, and $L_{2}=\hL_{z}$.  By the relation
$ [A,L_{1}]=[\hat{S}_{x},\hat{S}_{z}]=-i\hbar\hat{S}_{y},
$ we obtain the following bound for the noise.
\begin{equation}\label{eq:spin-bound}
\ep(\ps)^{2}
\ge\frac{\hbar^{2}\bracket{\hat{S}_{y}(t)}^{2}} {4[\Delta
\hat{S}_{z}(t)]^{2}+4[\Delta \hL_{z}(t)]^{2}}.
\end{equation} 
For apparatuses with large $[\Delta \hL_{z}(t)]^{2}$, the
optimal bound achieves when the numerator of the right-hand-side of
\Eq{spin-bound} is maximized.  This is achieved by $\psi=\al_{y}$, for
instance, in which we have $\langle\hS_{y}(t)\rangle=\Delta
\hS_{z}(t)={\hbar}/{2}$.
In this case, we have the optimal
bound as follows.
\begin{equation}\label{eq:3.4}
\ep^{2}\ge\ep(\al_{y})^{2}\ge
\frac{\hbar^{2}}{4+16[\Delta \hat{m}_{z}]^2},
\end{equation}
where $\hat{m}_{z}$ is the initial angular momentum normalized by
$\hbar$, i.e., $\hat{m}_{z}=\hL(t)_{z}/\hbar$. 
 If $\De \hat{m}_{z}$ is not large enough, the
right-hand-side of
\Eq{3.4} may not be optimal; however,
\Eq{3.4} still gives a correct lower bound, since our derivation uses no
approximation.   

For spin 1/2 measurements, the mean-square error is considered to
be the $\hbar^{2}$ times the error probability, and hence 
we should define the error probability $P_{e}(\ps)$ by 
\beq 
P_{e}(\ps)=\frac{\ep(\ps)^{2}}{\hbar^{2}}.
\eeq 
Then, the maximum error probability $P_{e}$ is bounded by
\beql{opt-error-prob} 
P_{e}\ge P_{e}(\al_{y})\ge\frac{1}{4+16[\Delta\hat{m}_{z}]^2}.
\eeq 
For the engineering of microscopic information processors such as
quantum logic gates \cite{NC00}, this bound is considered to be a serious
obstacle to realizing small and accurate quantum devices.  

In addition to the formulation discussed above, Yanase \cite{5} and
Wigner \cite{7} considered the measuring interaction with the following
form:
\begin{mathletters}
\begin{eqnarray}\label{eq:YWinteraction} U(\alpha_{x}\otimes \xi) 
&=& \alpha_{x}\otimes \xi^{+} + \beta_{x}\otimes \eta^{+},
                                                 \label{eq:A3a}\\ U(\beta_{x}\otimes \xi)
&=& \beta_{x}\otimes \xi^{-} + \alpha_{x}\otimes \eta^{-}.
                                                 \label{eq:A3b}
\end{eqnarray}
\end{mathletters}%
The states $\xi^{\pm}$ and $\et^{\pm}$  are not
normalized.   The states  $\xi^{\pm}$ are assumed to be eigenstates of the
observable
$M$ satisfying
\begin{eqnarray} M\xi^{\pm} &=&\pm  {\hbar\over 2}\xi^{\pm}.       
\label{eq:A4a}
\end{eqnarray} The problem is to find a lower bound of the sum of the
two ``unsuccessful probabilities'' $\|\eta^{+}\|^{2}$ and
$\|\eta^{-}\|^{2}$, 
\begin{equation}
\ep_{Y}^{2} = \|\eta^{+}\|^{2} + \|\eta^{-}\|^{2}, \label{eq:A4+1}
\end{equation} to show a trade-off with the \lq\lq size'' of the apparatus
characterized by the mean-square, 
$\bracket{\hat{m}_{z}^{2}}$, of the normalized angular momentum. 

Under these, and the additional technical assumptions that $\ep_{Y}$ is
very small and that $\bracket{\hat{m}_{z}^{2}}$ is so large that the
eigenvalues of $\hat{m}_{z}$ can be treated as a continuous parameter, 
Yanase \cite{5} obtained the following lower bound.
\begin{equation}\label{eq:AYanase}
\ep_{Y}^{2}>\frac{1}{8\bracket{\hat{m}_{z}^{2}}}.
\end{equation}
Later, Ghirardi et al.~\cite{6} derived the same bound for
rotationally invariant interactions without continuous parameter
approximation.

In what follows, we shall obtain a tighter bound for $\ep_{Y}^{2}$ 
without any approximation.  For this purpose, we shall
show the relation
\beq
\ep_{Y}^{2}\ge\frac{2}{\hbar^2}\ep(\al_{y})^{2}=2P_{e}(\al_{y}).                           
\label{eq:A3.5}
\eeq The proof runs as follows.  Easy computations show
\begin{mathletters}
\begin{eqnarray} UN(\alpha_{x}\otimes\xi)  &=& \beta_{x} \otimes(M -
\frac{\hbar}{2}I)\eta^{+},\label{eq:A3.5a}\\ UN(\beta_{x}\otimes\xi)
&=& \alpha_{x}\otimes(M + \frac{\hbar}{2}I)\eta^{-}.\label{eq:A3.5b}
\end{eqnarray}
\end{mathletters}%
By the relation 
$2\alpha_{y} = (1+i)\alpha_{x} + (1-i)\beta_{x}$, we have
\begin{eqnarray*}
\ep(\al_{y})^{2}
&=&
\|UN(\alpha_{y}\otimes\xi)\|^2\\ &=& \frac{1}{2}\|(M -
\frac{\hbar}{2}I)\eta^{+}\|^{2}       
   +\frac{1}{2}\|(M + \frac{\hbar}{2}I)\eta^{-}\|^{2}\\ 
&\le& 
\frac{\hbar^{2}}{2}\|\eta^{+}\|^{2}+\frac{\hbar^{2}}{2}\|\eta^{-}\|^{2}                      
=
\frac{\hbar^2}{2}\ep_{Y}^{2}.                         
\end{eqnarray*}
Thus, we obtain \Eq{A3.5}. By combining relations
(\ref{eq:opt-error-prob}) and (\ref{eq:A3.5}), we conclude
\begin{equation}\label{eq:AOzawa}
\ep_{Y}^{2}\ge\frac{1}{2+8(\De m_{z})^{2}}.  \label{eq:A3.6}
\end{equation} Under the conditions (i) $1\ll (\De m_x)^2$ and (ii)
$\langle
\hat{m}_z\rangle\approx 0$,  Yanase's bound, (\ref{eq:AYanase}),  turns
out to be a good approximation for the rigorous bound,
\eq{AOzawa}, and otherwise the new bound is tighter.

In order to show that the bound \eq{opt-error-prob}
typically vanishes for macroscopic apparatuses, we assume that the probe is
a three-dimensional isotropic harmonic oscillator in a coherent state. Let
$|\alpha\rangle$ and
$|\beta\rangle$ be the coherent states quantized along the $x$ and $y$
axes, respectively. Then from Ref. \cite{18} we have
\begin{equation} 
(\Delta \hat{m}_z)^2 = |\alpha|^2 + |\beta|^2,
\end{equation} 
and hence the optimal bound turns to 
\begin{equation}
P_{e}\ge\frac{1}{4+16|\alpha|^2+16|\beta|^2}.
\end{equation} 
If the probe is a macroscopic harmonic oscillator,  we
have $|\alpha|^2, |\beta|^2 \gg 1$, and hence the error probability
$P_{e}$ can be arbitrarily small. 

We have obtained a bound for the accuracy of general measurements
imposed from conservation laws and uncertainty relations.  This bound
shows that in order to make a precise measurement, the probe is required
to have very large variance of the conserved quantity, as long as the
probe can be observed repeatably.   If the apparatus is macroscopic, this
bound poses no serious limit.  However, for quantum information
processing, measuring interactions occur between qubits, which can
have only a small amount of conserved quantities.  
The relevance of this bound with the
fundamental limit of quantum information processing will be worth further
investigations.

\acknowledgments 
The author would like to thank Professor M. M.
Yanase and Professor A. Shimony for helpful discussions.


\begin{references}
\bibitem{2} E. P. Wigner, Z. Phys. {\bf 133}, 101 (1952).
\bibitem{3} H. Araki and M. M. Yanase, Phys. Rev. {\bf 120}, 622
(1960).
\bibitem{4} See also, H. Stein and A. Shimony, in {\it Foundations of 
Quantum Mechanics}, edited by D'{E}spagnat (Academic, New York,
1971).
\bibitem{5} M. M. Yanase, Phys. Rev. {\bf 123}, 666 (1961).
\bibitem{6} See also, G. C. Ghirardi, F. Miglietta and A. Rimini and  T.
Weber, Phys. Rev. D {\bf 24}, 347 (1981); {\bf 24}, 353 (1981).
\bibitem{7} E. P. Wigner, Am. J. Phys. {\bf 31}, 6 (1963).
\bibitem{8} T. Ohira and P. Pearle, Amer. J. Phys. {\bf 56}, 692 (1988).
\bibitem{indirect} M. Ozawa, Phys.\ Rev.\ A {\bf 62}, 062101 (2000);
{\bf 63}, 032109 (2001).
\bibitem{note1}  For a proof for unbiased measurements, see M. Ozawa,
in {\it Quantum Aspects of Optical Communications}, edited by C.
Bendjaballah et al., pp.~3--17,  Lecture Notes in Physics {\bf 378}
(Springer, Berlin,  1991).   A general proof will be published in a
forthcoming paper.
\bibitem{NC00} M. A. Nielsen and I . L. Chuang, Quantum Computation
and Quantum Information (Cambridge University Press, Cambridge,
2000). 
\bibitem{18} M. M. Nieto, Phys. Rev. Lett. {\bf 18}, 182 (1967).
\end{references}
\end{document}